\documentclass[aps,pre,preprint,groupedaddress,showpacs]{revtex4-1}
\usepackage[dvips]{graphicx}
\usepackage{textcomp}
\usepackage{amssymb}

\newcommand{\mc}{\multicolumn}
\begin{document}

\title{
\Large\bf A variance reduced estimator of the connected two-point function 
in the presence of a broken $\mathbb{Z}_2$-symmetry}

\author{Martin Hasenbusch}
\email[]{Martin.Hasenbusch@physik.hu-berlin.de}
\affiliation{
Institut f\"ur Physik, Humboldt-Universit\"at zu Berlin,
Newtonstr. 15, 12489 Berlin, Germany}

\date{\today}

\begin{abstract}
The exchange or geometric cluster algorithm allows us to define a variance 
reduced estimator of the connected two-point function in the presence of a
broken $\mathbb{Z}_2$-symmetry. We present first numerical tests for the
improved Blume-Capel model on the simple cubic lattice. We perform simulations
for the critical isotherm, the low temperature phase at vanishing external
field and, for comparison, also  the high temperature phase.
For the connected two-point function a substantial reduction of the variance 
can be obtained, allowing us to compute the correlation length $\xi$ with 
high precision. Based on these results, estimates for various universal
amplitude ratios that characterise the universality class of the 
three-dimensional Ising model are computed.

\end{abstract}

\maketitle

\section{Introduction}
Cluster algorithms \cite{SwWa87,Wolff} have drastically reduced auto-correlation
times in Monte-Carlo simulations of a certain class of spin models. 
In particular for the Ising model, critical slowing down could be virtually
eliminated. In addition, cluster algorithms
allow to introduce variance reduced estimators of the two-point function.
In the case of the Swendsen-Wang algorithm, after freezing or 
deleting links, the remaining degrees of freedom are the overall signs of 
the clusters. The variance reduced, or improved estimator is constructed by
performing the sum over these degrees of freedom exactly 
\cite{Wolff88,MH90,Wolff90}. This allowed to
determine the magnetic susceptibility and the correlation length of the 
Ising model and also $O(N)$-invariant non-linear $\sigma$-models with $N>1$
in the disordered phase to high precision. See for example refs. 
\cite{MHXY,MHamplitude}. However in the presence of a broken symmetry,
these estimators fail to reduce the variance considerably.

The exchange cluster algorithm \cite{ReMaCh98,ChMaRe98} is closely related with
the geometric cluster algorithm \cite{HeBl98}. In the exchange
cluster algorithm, a pair of systems is considered. These systems do not
interact. Hence the Hamiltonian of the pair is just given by the 
sum of the two Hamiltonians. In the exchange cluster algorithm, the values
of spins at corresponding sites are exchanged between the two systems. 
Since the total sum of the spins stays constant under such updates, the 
exchange cluster algorithm is not ergodic. Therefore, in addition, updates
of the individual systems with, for example, the local heat-bath and 
standard cluster
algorithms are performed. In the geometric cluster algorithm only a single
system is considered. The sites of the lattice are grouped into pairs. 
The values of the spins are exchanged within these pairs.
The authors of \cite{ReMaCh98,ChMaRe98} were mainly aiming at systems 
with external fields. Here the virtue of the algorithm is that 
the external field does not effect the  exchange of the spins. Therefore
in particular in the case of the Ising model in a random field one would
expect a reduction of auto-correlation times \cite{MaNeCh00}.

In \cite{mysphere,myfilm} we used the exchange cluster algorithm to get variance
reduced estimators of quantities related to the thermodynamic Casimir force.
Here, we discuss a variance reduced estimator of the connected
two-point correlation function in the presence of a broken $\mathbb{Z}_2$
symmetry.
We study the properties of this estimator at the example of the Blume-Capel 
model on the simple cubic lattice. Its reduced Hamiltonian is given by
\begin{equation}
\label{BlumeCapel}
H = -\beta \sum_{<xy>}  s_x s_y
  + D \sum_x s_x^2   - h \sum_x s_x  \;\; ,
\end{equation}
where the spin might assume the values $s_x \in \{-1, 0, 1 \}$.
$x=(x_0,x_1,x_2)$
denotes a site on the simple cubic lattice, where $x_i \in \{1,2,...,L_i\}$
and $<xy>$ denotes a pair of nearest neighbours on the lattice. We impose
periodic boundary conditions in all three directions. In our numerical 
study we consider lattices with the same linear extension $L=L_0=L_1=L_2$ in 
all directions.
The inverse temperature is given 
by $\beta=1/k_B T$, $D$ controls the density of vacancies $s_x=0$, and $h$ 
is an external field. One finds that
for $D^*=0.656(20)$ leading corrections to scaling vanish \cite{MHcritical}. 
Here we shall study the model
at $D=0.655$, where $\beta_c=0.387721735(25)$ is known with high precision 
\cite{MHcritical}.

The paper is organized as follows. First we recall the definition of the 
exchange cluster algorithm and discuss the construction of the variance
reduced estimator of the connected two-point function. Next we discuss the 
definition of the second-moment and the exponential correlation length.
We recall how these quantities are determined from the connected two-point 
function that we compute in the Monte Carlo simulation. Then we summarize
some results for critical phenomena which are needed for the analysis of 
our data.  Theoretical predictions for the behaviour of the  
slice-slice correlation function are summarized. It follows the discussion 
of our numerical study.  We briefly discuss the update scheme that is used.
The behaviour of the statistical error of the slice-slice correlation function
is analysed. Based on our data we study the critical behaviour in the 
high and the low temperature phase and on the critical isotherm. Here we
are mainly aiming at universal amplitude ratios. We  summarize
our results and given an outlook.
In the appendix we briefly summarize results that we obtained for the 
critical isotherm of the standard Ising model.

\section{The connected two-point function: variance reduction}
Let us start the discussion assuming $h>0$, such that the $\mathbb{Z}_2$ 
symmetry is explicitly broken.
The connected two-point function is defined  by
$ G(x-y) = \langle s_x s_y \rangle - \langle s_x \rangle \langle s_y \rangle$ ,
where $\langle s_x \rangle = \langle s_y \rangle =m$ is the magnetisation of 
the system.
Now let us consider a pair of identical systems. The two-point
function of the difference of the spins in these two systems is
\begin{eqnarray}
\label{G2}
 G_2(x-y) &=& \langle (s_{x,1} - s_{x,2})(s_{y,1}-s_{y,2}) \rangle \nonumber \\
          &=& \langle s_{x,1} s_{y,1} \rangle
            + \langle s_{x,2} s_{y,2} \rangle
            - \langle  s_{x,2} s_{y,1} \rangle - \langle  s_{x,1} s_{y,2} \rangle \;\;,
\end{eqnarray}
where the second index of $s_{x,l}$ with $l\in \{1,2\}$ denotes the system.
Since the two systems do not interact, 
$ \langle  s_{x,2} s_{y,1} \rangle = \langle  s_{x,2} \rangle \langle  s_{y,1} \rangle = m^2 $
and hence 
$G_2(x-y)  = 2 G(x-y)$  .

Now let us apply the exchange cluster algorithm to the pair of systems. 
The elementary operation of the algorithm is to swap the value of spins 
between the two systems. This can be written in terms of an auxiliary Ising 
variable $\sigma_{x} \in \{-1,1\}$:
\begin{equation}
\tilde s_{x,1} = \frac{1 +\sigma_x}{2} s_{x,1} + \frac{1 -\sigma_x}{2} s_{x,2}  \;\;,\;\;\;\;
\tilde s_{x,2} = \frac{1 -\sigma_x}{2} s_{x,1} + \frac{1 +\sigma_x}{2} s_{x,2}  \;.
\end{equation}
For $\sigma_x=-1$ the exchange is performed, while for $\sigma_x=1$ the
old values are kept.
Now we update the $\sigma_x$ using the Swendsen-Wang cluster algorithm.  
The construction of the clusters
is characterized by the probability to delete the link $<xy>$ between the 
nearest neighbours $x$ and $y$ \cite{HeBl98}:
\begin{equation}
\label{pdsimple}
 p_d = \mbox{min} [1, \exp(-2 \beta_{embed} )]  \;,
\end{equation}
where
$\beta_{embed} = \frac{\beta}{2} ( s_{x,1} - s_{x,2} )  ( s_{y,1} - s_{y,2} )$. 
A link $<xy>$  that is not deleted is called frozen. Clusters are sets of 
sites that are connected by frozen links. For all sites $x$ within
a given cluster $\sigma_x = \hat \sigma_{i}$, where $i$ labels the clusters.
Hence the remaining degrees of freedom are the $\hat \sigma_{i}=\pm 1$, with 
equal weight for each of the two possible values.
Variance reduced estimators are obtained by performing the average over all 
possible configurations of the $\hat \sigma_{i}$ exactly.  For the estimator 
$ A_2 = (s_{x,1} - s_{x,2})(s_{y,1}-s_{y,2}) $ 
we get the variance reduced counterpart
\begin{eqnarray}
\label{improved}
 A_{2,imp} &=& \frac{1}{2^{N_c} }
     \sum_{\hat \sigma}  [\sigma_x (s_{x,1} - s_{x,2})][\sigma_y (s_{y,1}-s_{y,2})]
         = \frac{1}{2^{N_c} } \sum_{\hat \sigma} \hat \sigma_{i | x \in i} 
                               \hat \sigma_{j | y \in j} (s_{x,1} - s_{x,2}) (s_{y,1} - s_{y,2}) 
       \nonumber \\
         &=& \Theta(x,y) \;  (s_{x,1} - s_{x,2}) (s_{y,1} - s_{y,2}) \;\;,
\end{eqnarray}
where $N_c$ is the number of clusters and $\Theta(x,y)$ is equal to $1$ if $x$ and $y$ belong 
to the same cluster and $0$ otherwise. Inspecting eq.~(\ref{pdsimple}) we 
see that $p_d<1$ requires that 
$(s_{x,1} - s_{x,2}) (s_{y,1} - s_{y,2}) > 0$. 
Hence the difference $(s_{x,1} - s_{x,2})$ has the same sign for all sites
$x$ in a given cluster. Hence $A_{2,imp} \ge 0$,
which is obviously not the case for the standard estimator $A_2$.

Next let us discuss the case of spontaneous symmetry breaking in the low 
temperature phase. The problem is, that for $h=0$ there is no symmetry
breaking on a finite lattice. In analytical calculations, one therefore 
introduces a finite external field $h$ and takes the thermodynamic
limit at finite $h$ first and then performs the limit $h \searrow 0$.  
In Monte Carlo simulations it is too cumbersome to mimic this approach. 
Therefore usually the magnetisation at $h=0$ is computed as
\begin{equation}
\label{magdef}
m = \frac{1}{L_0 L_1 L_2} 
\left \langle \left |\sum_x s_x \right| \right \rangle \;\;.
\end{equation}
This is motivated by the hypothesis that the partition function is dominated 
by configurations that can be clearly assigned to one of the bulk phases,
while the remainder is again dominated by configurations, where two interfaces
separate regions that can be assigned to the bulk phases. 
The contribution of the latter configurations is, at least in the most simple
approximation, proportional to $\exp(- 2 \sigma L^2)$, where $\sigma$ is the interface tension.
For a more detailed discussion see the vast literature on the physics of interfaces. See
for example ref. \cite{ScViBi14}  and references therein. 

In the same spirit, we align the magnetisation of the two systems here. 
To simplify the discussion, we ignore configurations with exactly vanishing
magnetisation in the following. First note that the constraint $M_1 M_2 > 0$,
where $M_l = \sum_x s_{x,l}$ does 
not affect the marginal distributions of the individual systems $l=1$ and $2$. 
Concerning the estimator of the two-point function, the discussion below 
eq.~(\ref{G2}) has to be slightly modified:
\begin{eqnarray} 
 \langle  s_{x,2} s_{y,1} \rangle = \langle  s_{x,1} s_{y,2} \rangle &=& 
\left \langle \frac{1}{L_0 L_1 L_2} \sum_u s_{u,1} \;\;
  \frac{1}{L_0 L_1 L_2} \sum_w s_{w,2} \right \rangle \nonumber \\
&=& \left \langle  \frac{1}{L_0 L_1 L_2} \left | \sum_u s_{u,1} \right | \;\;
  \frac{1}{L_0 L_1 L_2} \left |\sum_w s_{w,2} \right| \right \rangle 
= m^2 \;\;\;, 
\end{eqnarray}
where we used that the two systems are uncorrelated up to the constraint 
$M_1 M_2 > 0$. 

Now let us discuss how this constraint is imposed in the simulation.
Updating the individual systems by using local or cluster algorithms,
leaves the Boltzmann distributions of the individual systems invariant.
However, the resulting configurations  might violate the constraint 
$M_1 M_2 > 0$.
This could be reinforced by hand: If $M_1 M_2 < 0$ we simply multiply
all spins in the first system by $-1$. Since $M_2>0$ and $M_2<0$ are equal 
probable, this operation leaves invariant the Boltzmann distribution of the
first system. Now the aligned configurations are updated with the exchange
cluster algorithm and the improved estimator~(\ref{improved}) is computed.
The remaining problem is that the exchange cluster algorithm does not 
strictly leave the constraint $M_1 M_2 > 0$ invariant. By construction 
$M_1+M_2$ is kept constant. Based on the hypothesis on the probability
distribution of the magnetisation $P(M)$ discussed above,
the probability that $M_1 M_2$ changes sign under the exchange cluster 
algorithm is at least suppressed by a factor of $\exp(-2 \sigma L^2)$.

In our simulations, we actually considered the quantity
\begin{equation}
P = \sum_x s_{x,1} s_{x,2}  \;\;,
\end{equation}
which is invariant under the exchange of spins between the configurations.
We replaced the constraint $M_1 M_2 > 0$ by $P>0$. This means that after 
performing the updates of the individual systems, we determine $P$ and if 
$P<0$, the spins of the first system are multiplied by $-1$. The  
remaining question is, how likely is $M_1 M_2>0$ given $P>0$. In fact 
our numerical results show that with increasing $L$, the probability 
rapidly goes to one.

\section{The correlation length and the spectrum of the transfer matrix}
In our study we are aiming at the magnetic susceptibility and the correlation
length, which are derived from the two-point function. Here we briefly 
recall some basic definitions. For a more detailed discussion see  for 
example section 4 of ref. \cite{myhabil}.

In order to simplify the analysis, one projects to zero-momentum states of the 
transfer matrix. To this end one considers the correlation function
$\bar{G}(r) = \langle S_0 S_r \rangle - \langle S_0 \rangle \langle S_r \rangle$
of slices
\begin{equation}
\label{slice}
 S_{x_0} = \frac{1}{\sqrt{L_1 L_2}} \sum_{x_1,x_2} s_{(x_0,x_1,x_2)} \;.
\end{equation}
For finite $L_1$, $L_2$ and $L_0 \rightarrow \infty$, 
the slice-slice correlation function has the form
\begin{equation}
 \bar{G}(r) = \sum_{\alpha} c_{\alpha} \exp(-m_{\alpha} r) \;\;,
\end{equation} 
where 
\begin{equation}
c_{\alpha} = \langle 0 | S | \alpha \rangle^2 \;\;,
\end{equation}
where $| \alpha \rangle$ are the eigenvectors of the transfer matrix.
$| 0 \rangle$ is the eigenvector corresponding to the largest eigenvalue
$\lambda_0$. Since the transfer matrix is a real, symmetric and positive definite
matrix, the eigenvalues $\lambda_{\alpha}$ are real and positive. Let us assume that
they are ordered such that $\lambda_{\alpha} \ge \lambda_{\beta} $ for 
$\alpha < \beta$. The masses are given by 
$m_{\alpha} = - \ln(\lambda_{\alpha}/\lambda_0)$. In the basis of slice
configurations, $S$ is a diagonal matrix with entries given by eq.~(\ref{slice}).
The coefficient $c_{\alpha}$ is non-vanishing only if $| \alpha \rangle$
has zero momentum, zero angular momentum, and positive parity. 
For a more detailed discussion
of the transfer matrix formalism see for example section 4.1 of 
ref. \cite{myhabil}.
In the limit $L_1, L_2 \rightarrow \infty$ the dimension of the transfer 
matrix rapidly goes to infinity. One expects that
the time-slice correlation function assumes the form
\begin{equation}
\bar{G}(r) = \sum_i c_i \exp(-m_i r) + \sum_j f_{cut,j} (r) \;.
\end{equation}
In a particle interpretation, $m_1$ is the mass of the fundamental particle, 
while the $m_i$ with $i>1$ can be interpreted as masses 
of bound states of the fundamental particle.  The contributions 
\begin{equation}
f_{cut,j}(r) = \int_{\mu_{0,j}}^{\mu_{max,j}} \mbox{d} \mu \; a_j(\mu) 
\exp(-\mu r) \;
\end{equation}
are due to scattering states. Therefore $\mu_{0,j} = \sum_i n_{i,j} m_i$, 
where $n_{i,j}$ is integer and $\sum_i n_{i,j} > 1$.

The exponential correlation length is defined by the decay of the 
correlation function at large distances. Hence $\xi_{exp}=1/m_1$.
Analysing data obtained from Monte Carlo simulations one often 
considers the effective correlation length 
\begin{equation}
\label{xieff}
\xi_{eff}(r) =-1/\ln\left[\frac{\bar{G}(r+1/2)}{\bar{G}(r-1/2)} \right] \;\;.
\end{equation}
The exponential correlation length is obtained 
as $\xi_{exp} = \lim_{r \rightarrow \infty} \xi_{eff}(r)$.

The second moment correlation length is defined by
$ \xi_{2nd}^2 = \frac{\mu_2}{2 d \chi}$,
where $d=3$ in our case and the magnetic susceptibility can be written as
$ \chi= \sum_{r=-\infty}^{\infty} \bar{G}(r) $ and
$\mu_2 = d \sum_{r=-\infty}^{\infty} r^2 \bar{G}(r) $.
For a single exponential decay,
$\bar{G}(r) = \exp(-r/\xi_{exp})$, one gets
\begin{equation}
\label{finite_a}
 \xi_{2nd,single}^2 = \frac{\exp[-1/\xi_{exp}]}{(1-\exp[-1/\xi_{exp}])^2}
\end{equation}
for the second moment correlation length. In the limit
$\xi_{exp} \rightarrow \infty$ one gets
$\xi_{exp}/\xi_{2nd,single} = 1 + O(1/\xi_{exp}^2)$. For example, for
$\xi_{exp}=1$ we get $\xi_{exp}/\xi_{2nd,single} = 1.04219...$ . In order
to improve the convergence, we have multiplied $\xi_{exp}/\xi_{2nd}$ by
\begin{equation}
\label{finite_corr}
c_{a}=\xi_{2nd,single}/\xi_{exp}
\end{equation}
in our numerical analysis below.
Analysing our Monte Carlo data, we computed $\chi$ and $\mu_2$ in the
following way: Up to a certain distance $R$ we have used $\bar{G}(r)$
computed directly
from the configurations that we have generated. Since the relative
statistical error increases exponentially with the distance $r$, for $r>R$ we
have used instead
\begin{equation}
\label{extrapol}
\tilde G(r) = \bar{G}(R) \exp\left(-\frac{r-R}{\xi_{eff}(R+1/2)} \right) \;\;.
\end{equation}
We will comment on the choice of $R$ below.   Also note that,
in order to reduce the statistical error, 
we computed the slice-slice correlation function for all three directions
of the lattice. Furthermore, we exploited the translational invariance of the 
lattice.

\subsection{Results given in the literature}
The authors of ref.~\cite{pisa97} studied the behaviour of the 
correlation function in the high temperature phase of $O(N)$-invariant
models in three dimensions by using perturbation theory, high temperature
series expansions and the large $N$-expansion.  They conclude that 
the leading cut contribution is associated with a three particle state
with $\mu_1 = 3 m_1$. Furthermore, no bound state with 
a mass less than $3 m_1$ should contribute. They arrive at the estimate 
\begin{equation}
\lim_{t \searrow  0} \; \xi_{exp}/\xi_{2nd} =  1.000200(3)
\end{equation}
for the Ising universality class, where $t$ is the reduced temperature.

In the low temperature phase there should be a contribution from 
a cut characterized by $\mu_1 = 2 m_1$. It has been computed by the 
author of \cite{Pr98} at one loop level of perturbation theory. This
calculation was extended to two loop in \cite{CaHaPr99}. Corresponding
estimates are 
\begin{equation}
\label{frcut}
\lim_{t \nearrow  0} \; \xi_{exp}/\xi_{2nd} \approx  1.00652 \;\;\; (\mbox{at} \;\; 1-\mbox{loop}) 
\;\;\; \mbox{and} \;\;\; 
                      1.01266 \;\;\; (\mbox{at} \;\; 2-\mbox{loop})\;.
\end{equation}
In \cite{CaHaPr99} the correlation matrix of a large number of different
observables was determined in a Monte Carlo simulation of the Ising model
and the $\phi^4$ model on the simple cubic lattice. The analysis of these
data has shown that there is a bound state with 
\begin{equation}
\label{ourm2}
m_2 = 1.83(3)  \;\;.
\end{equation}
This result was confirmed by solving the Bethe-Salpeter equation for the
$\phi^4$ theory in three dimensions at one-loop level of 
perturbation theory \cite{CaHaPrZa02}.  
Correspondingly we \cite{MHamplitude} find that the ratio 
\begin{equation}
\label{fratiom}
\lim_{t \nearrow  0} \; \xi_{exp}/\xi_{2nd} = 1.020(5)
\end{equation}
is larger than the estimates~(\ref{frcut}) obtained from perturbation 
theory.

On the critical isotherm, for 
symmetry reason, we expect that, similar to the low temperature phase,
there is a cut characterized by $\mu_1 = 2 m_1$. Taking the 
numerical results for the linear lattice size 
$L=120$, given in table 1 of ref. \cite{EnFrSe03}, 
we get $\xi_{exp}/\xi_{2nd}=1.06(2)$.  
Note that the authors of ref. \cite{EnFrSe03} simulated
the improved $\phi^4$ model on the simple cubic lattice.
This result suggests that also for the critical isotherm there is a 
bound state with $m_2 < 2 m_1$.

\subsection{Analysing our numerical results}
Here we briefly summarize our preliminary study of $\bar{G}(r)$, 
which is the basis of our evaluation of the correlation length
below.

We fitted our numerical results for $\bar{G}(r)$ both in the low
temperature phase and for the critical isotherm with the Ansatz
\begin{equation}
 \bar{G}(r) = \sum_{i=1}^n c_i \exp(-m_i r)
\end{equation}
using $n=2$ and $3$.  In the case of the low temperature phase, we
find for all values of $\beta$ were we simulated at $m_2 \approx 1.8 m_1$, 
consistent with ref. \cite{CaHaPr99}. Furthermore $m_3 \approx 2.3 m_1$.
It is likely that this result is due to the cut at $2 m_1$.  Despite
the high statistical accuracy that we reached here for $\bar{G}(r)$, we where 
not able to get more precise results for the ratio $m_2/m_1$ than that
obtained in ref. \cite{CaHaPr99}, analysing the correlation matrix of several
observables. Therefore we shall not go into the details of our analysis.

For the critical isotherm, we find that $m_2 \approx 2.3 m_1$. The results
for $m_3$ depend very much on the range of $r$ that is fitted. We conclude
that there is no bound state with $m_2 < 2 m_1$. The main deviations from
a single exponential decay of $\bar{G}(r)$ are due to a cut with $\mu_1= 2 m_1$.

Below we shall use the effective correlation length to obtain our final
estimates of the exponential correlation length. We shall take the effective
correlation length at the distance $R = c \xi_{eff}$, selfconsistently.

In the high temperature phase, $\xi_{eff}$ very rapidly converges. We
take $c=2$, which should guarantee that  systematical errors are small
compared with the statistical ones.
In the case of the low temperature phase, we computed results for the two
choices $c=7$ and $9$.  In order to estimate the systematic error of our
result for the exponential correlation length, due to contributions of 
states with higher masses, we assumed $m_2=1.8 m_1$. Then, fitting with 
an Ansatz that contains two exponentials, we estimated the ratio of the 
two amplitudes. We obtained $c_2/c_1 \approx 0.04$ for the values of
$\beta$ we simulated at. Then, for this Ansatz, having inserted our
numerical estimate for the amplitude ratio, we computed $\xi_{eff}$. 
It turns our that the ratio $\xi_{exp}/\xi_{eff} \lessapprox $ $1.00012$ 
and $1.000024$ for $c=7$ and $9$, respectively. 

In the case of the critical isotherm we proceeded in a similar way, 
now assuming $m_2=2 m_1$. Based on our analysis we decided to take $c=6$,
where $\xi_{exp}/\xi_{eff} \lessapprox $ $1.00006$.

\section{Critical behaviour and universal amplitude ratios}
In this section we briefly summarize results needed for the analysis
of our numerical data. For a detailed discussion see for example the 
review  \cite{PeVi}. 
In the neighbourhood of the critical point various quantities 
diverge, following power laws. For example the exponential and 
the second moment correlation length at vanishing external field
behave as
\begin{equation}
\xi_{exp} \simeq f_{exp,\pm} |t|^{-\nu} \;\;\; , 
\;\;  \xi_{2nd} \simeq f_{2nd,\pm} |t|^{-\nu} \;,
\end{equation}
where $t=\beta_c-\beta$ is the reduced temperature. For simplicity 
we skip the usual normalization $1/\beta_c$. $f_{exp,\pm}$ and 
$f_{2nd,\pm}$ are the amplitudes and $\pm$ indicates whether the 
high ($+$) or the low temperature phase ($-$) is considered. The 
critical exponent of the correlation length $\nu$ is the same for all
systems in a given universality class. For a vanishing external 
field the magnetisation, the magnetic susceptibility and the specific
heat behave as
\begin{equation}
m \simeq B (-t)^{\beta}  \;\;\;,\;\; \chi \simeq C_{\pm} |t|^{-\gamma}
\;\;\;,\;\;  C_h \simeq A_{\pm} |t|^{-\alpha} \;.
\end{equation}
Note that here $\beta$ is, as usual, the critical exponent of the 
magnetisation. Also the behaviour on the critical isotherm, $\beta=\beta_c$
and $h \ne 0$, is given by power laws. In the following we assume $h>0$. 
The exponential and the second moment correlation length behave as
\begin{equation}
\xi_{exp} \simeq f_{exp,c} h^{-\nu_c}  \;\;\;,\;\; 
\xi_{2nd} \simeq f_{2nd,c} h^{-\nu_c} \;\;.
\end{equation}
The magnetisation and the magnetic susceptibility behave as
\begin{equation}
m \simeq B_c h^{1/\delta} \;\;\;,\;\; \chi \simeq C_c h^{1/\delta-1} \;\;.
\end{equation}

The critical exponents $\nu$, $\beta$, $\gamma$, $\alpha$, $\nu_c$ 
and $\delta$ are the same for all systems in a given universality 
class, which is in our case the universality of the Ising model in 
three dimensions. Following renormalization group theory, the exponents
listed above can be expressed in terms of only two exponents.
For example one could express them in terms of the so called RG-exponents
$y_t$ and $y_h$, where the subscript $t$ indicates a thermal perturbation
and $h$ a perturbation by the external field:
\begin{equation}
\label{allexponents1}
 \nu = 1/y_t \; , \;\;
 \alpha = 2 - \frac{d}{y_t}  \; , \;\;
 \eta = d + 2 -2 y_h \; , \;\;
 \beta = \frac{d -y_h}{y_t} \; , \;\;
 \gamma = \frac{2 y_h - d}{y_t} \; , \;\;
\end{equation}
and for the critical isotherm
\begin{equation}
\label{allexponents2}
\nu_c = 1/y_h \; , \;\;
 \delta = \frac{y_h}{d-y_h} \;\;\;,
\end{equation}
where $d$ is the dimension of the system. 
Quite recently Simmons-Duffin \cite{SiDu15} computed  
the dimensions of the fields by using the conformal bootstrap with 
high precision
\begin{equation}
\label{preciseE}
3 - y_h = \Delta_{\sigma} = 0.518151(6) \;\;\;,\;\; 3 - y_t = \Delta_{\epsilon} = 1.41264(6) \;\;\;.
\end{equation}
These results are fully consistent with, but clearly more accurate than
\begin{equation} 
 \nu = 0.63002(10)  \;\;\; , \;\; \eta = 0.03627(10) \;\;\;
\end{equation}
obtained by a finite size scaling analysis of Monte-Carlo data obtained for 
the improved Blume-Capel model \cite{MHcritical}. For a comparison with the
vast amount of results obtained by various methods see \cite{SiDu15,MHcritical}.
Taking the results of \cite{SiDu15} one arrives at 
$\nu=0.629977(24)$, $\eta=0.036302(12)$, $\gamma=1.237084(54)$, 
$\beta=0.326423(16)$, $\alpha=0.110069(71)$, 
$\nu_c=0.4029254(10)$, $\delta=4.789818(67)$, and
$1/\delta=0.208776(3)$. 

The $\simeq$ in the power laws listed above means that they are 
strictly valid only in the scaling limit $t \rightarrow 0$.  At finite
$t$ corrections have to be taken into account. For example the 
magnetic susceptibility behaves as 
\begin{equation}
\chi = C_{\pm} |t|^{-\gamma} \; \left(1+ a_{\pm} |t|^{-\theta}  + b t + 
c_{\pm} |t|^{-\theta'} + ... \right) + d(t) 
\end{equation}
where $d(t)$ is the analytic background. The terms $a_{\pm} |t|^{-\theta}$ and
$c_{\pm} |t|^{-\theta'}$ are singular or confluent corrections, 
while $b t$ is an analytic or non-confluent correction.  
Furthermore $\theta=\nu \omega$ and $\theta'=\nu \omega'$.  Various methods,
e.g. the $\epsilon$-expansion, perturbation theory in three dimensions fixed, 
high temperature series expansion and Monte-Carlo simulations of lattice 
models give consistently $\omega \approx 0.8$ for the exponent of the leading
correction. For the analysis of our data we shall use $\omega=0.832(6)$ 
\cite{MHcritical}.
The authors of \cite{ElShowketal14} obtained 
$\omega= \Delta_{\epsilon'}-3 = 0.8303(18)$ which slightly differs from
our central value. Note that in the case of the model studied here, the 
amplitude of leading corrections is small. Hence the precise value of 
$\omega$ has little influence on our final results.

There is a subleading correction due to the breaking of the
Galilean invariance of space by the simple cubic lattice. The associate
correction exponent is $\omega'' \approx 2$. For a precise estimate see 
\cite{pisa97}.  

Using the scaling field method, the authors of ref. \cite{NewmanRiedel}  find
a subleading correction with the exponent $\omega' = 1.67(11)$. Up to now, 
there is no confirmation of this finding by using other methods. In the 
following numerical analysis we shall assume the existence of this correction, 
which has little influence on central values, but enlarges the estimate
of systematic errors.

Concerning physics results we are mainly aiming at so called universal
amplitude ratios that are characteristic for the universality class
of the three-dimensional Ising model.
While individual amplitudes depend on the microscopic details of the 
model, certain combinations are universal. The combinations of the 
corresponding quantities have a critical exponent that is equal to zero
which means that they are renormalization group invariant or 
dimensionless. First we compute the ratios of amplitudes 
$f_{exp,+}/f_{2nd,+}$, $f_{exp,-}/f_{2nd,-}$, and $f_{exp,c}/f_{2nd,c}$.
The ratios $f_{2nd,+}/f_{2nd,-}$ and $C_{+}/C_{-}$ relate the low and 
high temperature phase.
For a broken $\mathbb{Z}_2$ symmetry we define the coupling 
\begin{equation}
\label{coupling}
u = \frac{3 \chi}{\xi_{2nd}^3 m^2} \;\;.
\end{equation}
For $h=0$, in the low temperature phase we get in the scaling limit
\begin{equation}
u^* = \lim_{t \nearrow 0} u(t,0) =  \frac{3 C_-}{f_{2nd,-}^3 B^2}
\end{equation}
and analogously
\begin{equation}
u_c =  \lim_{h \searrow 0} u(0,h) = \frac{3 C_c}{f_{2nd,c}^3 B_c^2}  
\end{equation}
for the critical isotherm. The quantity
\begin{equation}
 Q_2 = (f_{2nd,c}/f_{2nd,+})^{2-\eta}  \; C_+/C_c
\end{equation}
connects the critical isotherm with the high temperature phase.  Finally
\begin{equation}
 R_{\chi} = C_+ D_c B^{\delta-1} \;,
\end{equation}
where $h \simeq D_c m^{\delta}$, relates the critical isotherm with both
the low and the high temperature phase.

\section{The simulations}
The exchange cluster algorithm is not ergodic on it own. Therefore additional
updates of the individual systems are performed. In particular an update
cycle is composed of: 
\begin{itemize}
\item
One sweep with the local heat bath algorithm for both systems
\item
Standard cluster updates of both systems
\item
One sweep with the local Todo-Suwa \cite{ToSu13,Gutsch} algorithm for 
both systems
\item
One Swendsen-Wang exchange cluster update
\item
Random translation of one system
\end{itemize}
For lack of time, we did not optimize this update cycle. Let us briefly
discuss the choice of the cluster updates of the individual systems:
In the low temperature phase, we updated the individual systems by using
the single cluster algorithm. The number of single cluster
updates was chosen roughly as the total volume of the lattice  divided
by the average size of a cluster.
In the high temperature phase, we updated the individual systems by using 
the Swendsen-Wang algorithm. This allowed us to compare the variance 
reduced estimators of the correlation function that are based on the 
standard Swendsen-Wang cluster algorithm and the Swendsen-Wang version 
of the exchange cluster algorithm.

In the case of the critical isotherm, the cluster algorithm applied to 
the individual systems has to be modified to take the external field into
account \cite{Wang89,LaRi89}.
The construction of the clusters is the same as for a vanishing external 
field $h=0$.
Following ref. \cite{Wang89}, there are two ways to incorporate the external 
field.
The first one is by representing the external field by a ``ghost-spin''.
The link of a spin $s_x$ with the ghost-spin is frozen with the probability
\begin{equation}
p_{f,h}=1-p_{d,h} \;,
\end{equation}
where the delete probability $p_{d,h}=$min$[1,\exp(-2 h s_x)]$. All clusters
that are frozen to the ghost-spin keep the old sign of the spins. A cluster 
is frozen to the ghost-spin if it contains at least one spin that is frozen to 
the ghost-spin. Clusters that are not frozen to the ghost-spin get the sign
plus or minus with equal probability.

The alternative is to chose the new sign of the clusters with the heat-bath
probability 
\begin{equation}
p_C(-)=\frac{\exp\left(-h \sum_{x \in C} s_x\right)}
       {\exp\left(-h \sum_{x \in C} s_x\right)+\exp\left(h \sum_{x \in C} s_x\right)}
\end{equation}
and $p_C(+)=1-p_C(-)$.

Here we used a modified version of the ghost-spin variant. First we run through
all sites of the lattice and decide whether the spin is frozen to the ghost-spin
or not. Then we construct all clusters that contain spins that are frozen to the 
ghost-spin. As in ref. \cite{Wang89}, these clusters keep their sign. In contrast 
to \cite{Wang89}, we change the sign of all clusters that are not frozen to
the ghost-spin. This has the technical advantage that we need not construct these
clusters, since we just have to change the sign of all spins that do not 
belong to clusters that are frozen to the ghost-spin. A preliminary study 
shows that also auto-correlation times compare favourably.  In our update
cycle, we performed one of these updates for each system.

We used the SIMD-oriented Fast Mersenne Twister 
algorithm \cite{twister} as pseudo-random number generator. In total, 
all our simulations took about 18 years of CPU time on a single core 
of an Intel(R) Xeon(R) E5-2660 v3 running at 2.60GHz.

\subsection{The critical isotherm}
We simulated at the estimate of the inverse critical temperature $\beta=0.387721735$ at
various values of the external field. Preliminary simulations indicate that the deviation 
from the thermodynamic limit for the quantities that we study are below our statistical 
accuracy for $L \gtrapprox 11 \xi$.  Since the variance reduced quantities studied here
are self-averaging, we decided to simulate much larger lattices. Our final results are 
taken from simulations with $L \approx 40  \xi$.   Our results are summarized  in tables \ref{finiteh} and \ref{finiteh2}.

\begin{table}
\caption{\sl \label{finiteh}
Results for the critical isotherm $\beta=0.387721735$ of the Blume-Capel model 
at $D=0.655$.  In the first column we
give the value of the external field $h$. The second column contains the linear 
lattice size $L$. Next we give the number of update cycles divided by $10^5$. 
It follows the magnetic susceptibility, computed by using the improved
estimator. Then we give the results of the second moment correlation length
$\xi_{2nd}$ and the exponential correlation length $\xi_{exp}$. Next
we give the ratio $\xi_{exp}/\xi_{2nd}$, which is corrected by the factor
$c_a$, eq.~(\ref{finite_corr}), in the last column. 
All estimates given here are computed for $R = 6 \xi_{eff}$, 
eq.~(\ref{extrapol}).
}
\begin{center}
\begin{tabular}{lrclllll}
\hline
\mc{1}{c}{$h$} &  \mc{1}{c}{$L$}  & \mc{1}{c}{stat$/10^5$} & 
\mc{1}{c}{$\chi$} & \mc{1}{c}{$\xi_{2nd}$} &  \mc{1}{c}{$\xi_{exp}$}  & 
\mc{1}{c}{$\xi_{exp}/\xi_{2nd}$} & 
\mc{1}{c}{$c_a$ $\xi_{exp}/\xi_{2nd}$}  \\
\hline
0.02 &  60 & 100  & \phantom{00}4.65293(10) & \phantom{0}1.467172(38) &  \phantom{0}1.50849(16) &  1.02816(9) &  1.00957(9) \\ 
0.01  &  80 & 100  & \phantom{00}8.13646(17) & \phantom{0}1.948550(48) &  \phantom{0}1.98996(21) &  1.02125(9) &  1.01058(9) \\
0.006 & 100 &100  &\phantom{0}12.24571(26) & \phantom{0}2.398411(59) &  \phantom{0}2.44138(25) &  1.01792(9) &  1.01083(9) \\
0.003 & 130 &100  &  \phantom{0}21.27400(46) & \phantom{0}3.176033(80) &  \phantom{0}3.22465(34) &  1.01531(9) &  1.01125(9) \\
0.001 & 200 & \phantom{0}60 & \phantom{0}50.8996(15)  & \phantom{0}4.95094(17)  &  \phantom{0}5.01644(69) &  1.01323(12)&  1.01155(12) \\
0.0006& 248 & \phantom{0}40 &  \phantom{0}76.3007(27)  & \phantom{0}6.08355(24)  &  \phantom{0}6.1609(10)  &  1.01272(14)&  1.01161(14) \\
0.0002 &380 & \phantom{0}19 & 182.140(10)   & \phantom{0}9.47362(57)  &  \phantom{0}9.5850(23)  &  1.01176(20)&  1.01130(20) \\
0.0001 &500 & \phantom{0}16 & 315.267(20)   &12.52682(82)  & 12.6725(34)  &  1.01163(23)&  1.01136(23) \\
\hline
\end{tabular}
\end{center}
\end{table}

\begin{table}
\caption{\sl \label{finiteh2}
Further results for the critical isotherm $\beta=0.387721735$. 
In the first, second and third column we give the value of the external field $h$,
our results for the magnetisation $m$ and the renormalization group invariant quantity 
$u$, eq.~(\ref{coupling}), respectively.
}
\begin{center}
\begin{tabular}{lll}
\hline
\mc{1}{c}{$h$} &   \mc{1}{c}{$m$} &
\mc{1}{c}{$u$} \\
\hline
0.02  &  0.4543898(13) & 21.4066(13) \\
0.01  &  0.3939990(12) & 21.2536(13) \\
0.006 &  0.3544806(11) & 21.1909(12) \\
0.003 &  0.3069654(11) & 21.1415(13) \\
0.001 &  0.2442012(14) & 21.0997(17) \\
0.0006&  0.2195328(16) & 21.0950(20) \\
0.0002&  0.1745692(21) & 21.0884(30) \\
0.0001&  0.1510557(23) & 21.0864(32) \\
\hline
\end{tabular}
\end{center}
\end{table}

First let us discuss the performance of the improved estimator of the 
two-point function. Actually we did not directly determine the variance
of the quantities. During the simulation we computed the averages over bins
of 1000 measurements each. Hence we had only access to the statistical error
and not to the variance and the auto-correlation times separately.

Analysing the data for the standard estimator of the slice-slice correlation
function we find that the statistical error depends little on the distance
between the slices. Hence for the connected  slice-slice correlation
function the relative statistical error increases proportional to 
$\exp(r/\xi_{exp})$. The same holds for the effective correlation length
$\xi_{eff}$
computed from the  standard estimator of the slice-slice correlation
function.
On the contrary we find for all values of the external field $h$  
that the statistical error of the variance reduced estimator of the 
slice-slice correlation function decreases as $\exp(-r/[2 \xi_{exp}])$.
Hence the  relative statistical error increases as $\exp(r/[2 \xi_{exp}])$.
The same holds for the effective correlation length $\xi_{eff}$
computed from the  variance reduced estimator of the slice-slice
correlation function.

Now let us turn to the analysis of our data. First we fitted our data for 
the second moment correlation length, the  magnetisation, and the magnetic 
susceptibility using power law Ans\"atze. Then we studied universal
ratios that consist of quantities defined on the critical isotherm only.

We fitted the second moment correlation length with the Ans\"atze
\begin{equation}
 \xi_{2nd} = f_{2nd,c} h^{-\nu_c} \;\;\left( 1 + \sum_i^n a_i h^{\epsilon_i} 
\right) \;\; ,
\end{equation}
where $f_{2nd,c}$ and the $a_i$ are the free parameters of the fit.
We performed fits for $n=1$, $2$ and $3$, using different choices for the 
correction exponents $\epsilon_i$. As values we have used 
$\epsilon_i=0.832 \nu_c$, $1.67 \nu_c$, $2 \nu_c$, and $2$, with
$\nu_c=0.4029254$. 
It turns out that only for the exponent $\epsilon=2 \nu_c$  we find 
an amplitude that is clearly different from zero. In particular, fitting
the data with a single correction term and $\epsilon_1=2 \nu_c$ we find 
$f_{2nd,c} = 0.306321(17)$, $a_1=-0.161(18)$  and $\chi^2/$d.o.f. $=0.12$.
Our final estimate, and in particular  the error bar, is chosen such that 
the results of various plausible fits are accommodated. In order to obtain 
the dependence of the central value on $\nu_c$ and $\beta_c$ we repeated 
a selection of fits for slightly shifted values of $\nu_c$ and $\beta_c$.
We arrive at
\begin{equation}
 f_{2nd,c} = 0.30631(18) - 220 \; (\beta_c -0.387721735) \;\; 
                         -   3 \; (\nu_c - 0.4029254)  \;\;.
\end{equation}

Next we fitted the magnetisation with Ans\"atze 
\begin{equation}
 m = B_c \; h^{1/\delta} \;\;\left( 1 + \sum_i^n a_i h^{\epsilon_i} \right)
\end{equation}
using $n=1$ and $2$. 
In turns out that for $n=1$ and $\epsilon_1=2 \nu_c$, we get 
$\chi^2/$d.o.f. $=0.68$ taking all our values of $h$ into account.
One gets $B_c=1.03340069(28)$ and $a_1=-0.11479(10)$.  In order to 
get an estimate of possible systematic errors due to further corrections,
we performed fits with $n=2$, adding a term with a correction exponent 
$\epsilon_2= 0.832 \nu_c$ or $\epsilon_2= 1.67 \nu_c$.  In both cases, 
the amplitudes of the corresponding corrections remain compatible 
with zero within the error bars. In particular for the fit with 
$\epsilon_2=0.832 \nu_c$, the statistical error of $B_c$ increases considerably
compared with $n=1$ and $\epsilon_1=2 \nu_c$. 
We quote 
\begin{equation}
B_c = 1.033401(20) 
+ 170 \;\;(\beta_c - 0.387721735) + 7 \;\; (1/\delta-0.208776)
\end{equation}
as our final result. Next we have analysed the magnetic susceptibility.
Also here we find that all data can be fitted well with an Ansatz that 
contains a single correction term with the correction exponent 
$\epsilon=2 \nu_c$.  In particular we find 
$\chi^2/$d.o.f. $=0.89$ and $C_c= 0.2157487(34)$ and $a_1=-0.55805(66)$.
As in the case of the magnetisation we performed fits, where we added
a second correction term.
We arrive at our final estimate
\begin{equation}
C_c = 0.215749(15) + 73 \;\;(\beta_c - 0.387721735) 
+ 1.5  \;\; (1/\delta-0.208776) \;\;.
\end{equation}
The amplitudes of the magnetisation and the magnetic susceptibility on the 
critical isotherm are trivially related by $C_c = B_c /\delta$. Our numerical
estimates of $C_c$ and $B_c$ are indeed consistent with this relation.

Next we analysed the renormalization group invariant quantity $u$, 
eq.~(\ref{coupling}). We used the Ansatz
\begin{equation}
\label{ucansatz}
 u = u_c + c_1  \xi_{2nd}^{-\epsilon_1} + c_2  \xi_{2nd}^{-\epsilon_2} \;\;,
\end{equation}
where $u_c$, $c_1$ and $c_2$ are the free parameters.
We performed fits using $\epsilon_1=0.832$, which is our 
estimate of $\omega$ and the two choices $\epsilon_2=2 \omega$ and
$\epsilon_2=2$.

For both choices we get an acceptable $\chi^2/$d.o.f. 
taking into account all data except for our largest 
value of $h$. As our final result we take
\begin{equation}
\label{finaluc}
u_c = 21.086(20)  \;,
\end{equation}
which is the  value of $u$ for our smallest value of $h$.
The error bar is taken such that the results of  the 
fits discussed above are covered.
Since there is little variation of $c_a \xi_{exp}/\xi_{2nd}$
with $h$, we abstain from fitting our data. We just take the result obtained
for our smallest value of $h$ as estimate of the scaling limit
\begin{equation}
f_{exp,c}/f_{2nd,c} = 1.0114(4) \;.
\end{equation}
The error is chosen such that all results for $h \le 0.003$ are covered.

\subsection{The low temperature phase, $h=0$}
\label{Slowtemp}
First we studied finite size effects at $\beta=0.391$ and $\beta=0.42$ by simulating
a large range of lattice sizes.  Throughout we performed $10^8$ update cycles.
At the level of our statistical accuracy, the results for the correlation length, 
the magnetic susceptibility, the magnetisation, and the energy density are consistent among each 
other for $L \ge 12$ and $48$ for $\beta=0.42$ and $0.391$, respectively. 
Taking our final results $\xi_{exp} =1.08701(35)$ and  $4.4449(19)$, discussed below, 
we find consistently that for $L \gtrapprox 11 \xi_{exp}$ deviations from 
the thermodynamic limit are small compared with our statistical errors. 
Furthermore, the numerical results are consistent with an approach of the 
thermodynamic limit that is exponentially fast in the linear lattice size.
Concerning the validity of the variance reduced estimator, we checked 
whether the signs of
$P$ and $M_1 M_2$ coincide. For $\beta=0.42$  we find that this is the case
for the fraction $0.980740(46)$, $0.998769(12)$, and $0.9999823(14)$ of pairs
of configurations for the linear lattice sizes  
$L=4$, $6$, and $8$, respectively.
For the larger lattice sizes $L=10$, $12$, $16$, ... that we simulated, the sign of
$P$ and $M_1 M_2$ coincides for all configurations that we analysed.
For $\beta=0.391$, we find a fraction of $0.999972(18)$ for $L=32$, while for all 
larger lattice sizes that we simulated, the sign of $P$ and $M_1 M_2$ coincides for all 
configurations that we analysed. Furthermore  the analysis of our data
shows that the variance reduced estimator of the correlation function is self-averaging. 

Our final estimates are obtained for lattice sizes $L \gtrapprox 44 \xi_{exp}$, where deviations
from the thermodynamic limit a far smaller than our statistical errors. 
Our numerical estimates are summarized in table \ref{lowtemp}.

Similar to the critical isotherm we did not analyse autocorrelation times and variance 
separately. Instead we computed the statistical error using a jackknife analysis.
For the standard estimator of the slice-slice correlation  function $\bar{G}(r)$ we 
find that the statistical error virtually does not depend on the distance $r$.
Hence the signal to error ratio decreases as $\exp(-r/\xi_{exp})$.  In contrast, 
for the variance reduced estimator we find that the statistical error decreases 
as $\exp(-r/[2 \xi_{exp}])$. Hence the signal to error ratio decreases as
$\exp(-r/[2 \xi_{exp}])$.  Similar observations hold for the effective correlation
length $\xi_{eff}(r)$, which is computed from $\bar{G}(r)$. This  
improvement allowed us to take $\xi_{eff}(r)$ at about twice the distance
compared with ref. \cite{MHamplitude} as estimate of $\xi_{exp}$, making
systematical errors negligible.

\begin{table}
\caption{\sl \label{lowtemp}
Results for the low temperature phase of the Blume-Capel model at $D=0.655$
and a vanishing external field $h=0$.  
In the first column we give the inverse temperature $\beta$. In the remaining 
columns we give results for the same quantities as in table \ref{finiteh}
for the critical isotherm. Here we skip the ratio $\xi_{exp}/\xi_{2nd}$ and 
give only the corrected one $c_a \xi_{exp}/\xi_{2nd}$.
All estimates given here are computed for $R=7 \xi_{eff}$, eq.~(\ref{extrapol}).
}
\begin{center}
\begin{tabular}{lrclllll}
\hline
\mc{1}{c}{$\beta$} &  \mc{1}{c}{$L$}  &  \mc{1}{c}{stat$/10^5$} &
\mc{1}{c}{$\chi$} & \mc{1}{c}{$\xi_{2nd}$} &  \mc{1}{c}{$\xi_{exp}$}  &
\mc{1}{c}{$c_a$ $\xi_{exp}/\xi_{2nd}$} & \mc{1}{c}{$u$} \\
\hline
0.42  & \phantom{0}48 & 205 &  \phantom{00}1.964992(32) & 
\phantom{0}1.031143(20) & \phantom{0}1.08696(14) & 1.01786(12) &
14.07096(67) \\
0.41 &\phantom{0}60 &126 & \phantom{00}3.193168(68) &\phantom{0}1.302022(33)  &
  \phantom{0}1.35761(21) & 1.01949(14) & 14.06315(84) \\ 
0.40 & \phantom{0}88& 100  & \phantom{00}6.84022(17) & \phantom{0}1.892267(52)  & \phantom{0}1.95348(33) & 1.02116(16) & 14.06400(91)  \\
0.396 & 112 &104 & \phantom{0}11.24176(28) & \phantom{0}2.423518(68)  & \phantom{0}2.49420(41) & 1.02230(15) & 14.06351(91)  \\
0.394 & 132 & 100&  \phantom{0}15.89635(42) & \phantom{0}2.883053(83)  & \phantom{0}2.96210(51) & 1.02256(16) & 14.06513(93)  \\
0.393 & 148 &100& \phantom{0}19.74238(52) & \phantom{0}3.214815(90)  & \phantom{0}3.30138(55) & 1.02301(16) & 14.06754(91) \\
0.392 & 168 & 102& \phantom{0}25.65302(69) & \phantom{0}3.66821(11)   & \phantom{0}3.76268(63) & 1.02274(16) & 14.07048(93) \\
0.391 & 196 &101& \phantom{0}35.73272(97) & \phantom{0}4.33667(13)   & \phantom{0}4.44557(75) & 1.02295(16) & 14.07045(93) \\
0.39 &248 &100 &  \phantom{0}56.1525(15)  & \phantom{0}5.45174(16)   & \phantom{0}5.58651(95) & 1.02335(16) & 14.07095(92) \\
0.389 &400 &\phantom{0}33 & 114.9821(48)  & \phantom{0}7.84194(34)   & \phantom{0}8.0305(20)  & 1.02338(23) & 14.0755(14) \\
0.3883& 580 & \phantom{0}10 & 307.102(26)   &12.9205(12)    &13.2266(72)  & 1.02345(51) & 14.0771(29) \\
\hline
\end{tabular}
\end{center}
\end{table}

We analysed the data for the magnetisation obtained here along with those of 
ref. \cite{MHamplitude} by using Ans\"atze of the type
\begin{equation}
 m = B \; (-t)^{\beta} \; \;\;\left( 1 + \sum_i^n a_i \; (-t)^{\epsilon_i} \right)
\end{equation}
with $t=\beta_c-\beta$. We performed fits for $n=2$ and $3$. We fixed the exponents
$\beta=0.326423$, $\epsilon_1=0.832 \; \nu$, $\epsilon_2=1$ and $\epsilon_3=2 \;\nu$, where $\nu=0.629977$.
For $n=3$ we get $\chi^2/$d.o.f. close to one up to $\beta_{max}=0.41$, where we take all data for $\beta \leq \beta_{max}$
into account.
For fits with $n=2$ we get $\chi^2/$d.o.f. up to about $\beta_{max} = 0.395$.
Comparing the results of different fits, we arrive at the final estimate
\begin{equation}
 B = 1.9875(3)  + 2460 \;\;(\beta_c-0.387721735) \; + 22  \;\;(\beta - 0.326423) \;\;.
\end{equation}

In the case of the coupling $u$, we abstain from fitting, since there is little
variation with $\beta$. As final estimate we take the value obtained for our 
smallest value of $\beta$
\begin{equation}
 u^* = 14.08(1) \;\;.
\end{equation}
The error bar is chosen such that also the results for $\beta=0.389$ and
$0.39$ are covered. This result is fully consistent, but more precise
than our previous estimate $u^* = 14.08(5)$ \cite{MHamplitude}. For a 
comparison with results obtained by using other methods and previous 
Monte Carlo simulations see ref. \cite{MHamplitude}.

Also in the case of $c_a \xi_{exp}/\xi_{2nd}$ we abstain from fitting.
As our final estimate we take
\begin{equation}
\label{finalxx}
f_{exp,-}/f_{2nd,-}=1.0234(6) \;\;,
\end{equation}
where the error bar is chosen such that the results for our four 
smallest values of $\beta$ are covered. For all values of
$\beta$ we compared our result for $c_a \xi_{exp}/\xi_{2nd}$ using 
$R=7$ and $R=9$, eq.~(\ref{extrapol}). We conclude that the difference
should be clearly smaller than the error bar given in eq.~(\ref{finalxx}).
Our present result is consistent with but more precise than 
$f_{exp,-}/f_{2nd,-}=1.020(5)$ obtained in ref. \cite{MHamplitude}. For
a comparison with results obtained by using other methods and previous
Monte Carlo simulations see ref. \cite{MHamplitude}.

\subsection{The high temperature phase}
Finally we also performed simulations in the high temperature
phase. We simulated at values of the inverse temperature
$\beta_h = 2 \beta_c - \beta_l$, where $\beta_c=0.387721735$ is 
our estimate of the inverse critical temperature and $\beta_l$ are the 
values of $\beta$ that are used in section \ref{Slowtemp}. The linear 
lattices sizes $L$ are essentially the same as for the corresponding 
values of $\beta$ in the low temperature phase. 

We computed variance reduced estimators
both based on the standard Swendsen-Wang update and the cluster exchange
update of the two systems. In the following we use the subscripts $SW$ and 
$EC$ to discriminate between the two.
We find that the ratio of the statistical errors of $G_{EC}(r)$ and $G_{SW}(r)$ 
depends little on $r$. In both cases we find that the ratio of signal to 
statistical error decreases as $\exp(-r/[2 \xi_{exp}])$. The same holds for
the effective correlation length obtained from  $G_{SW}(r)$ and $G_{EC}(r)$.
For large distances, we find for all values of $\beta$ that we simulated
a ratio of about $1.55$ between the statistical errors of $G_{EC}(r)$ and 
$G_{SW}(r)$. For small distances we see a smaller factor that depends 
slightly on $\beta$. For our smallest $\beta$ we find a factor of about 
$1.5$ that decreases to about $1.2$ for our largest value of $\beta$. 
Note that in the case of  $G_{SW}(r)$ the measurements of both systems enters.
Hence the performance of the two variance reduced estimators is very 
similar. Results for various quantities derived from $G_{SW}(r)$ are summarized 
in table \ref{highresults}. 

\begin{table}
\caption{\sl \label{highresults}
Results for the high temperature phase of the Blume-Capel model at $D=0.655$. 
We give results for the same quantities as in table \ref{lowtemp}
for the low temperature phase. Only $u$ is missing, since it is not 
defined for a vanishing magnetisation.
All estimates given here are computed for $R \approx 2 \xi_{eff}$,
eq.~(\ref{extrapol}). The numbers are obtained from improved estimators based
on the standard Swendsen-Wang algorithm.
}
\begin{center}
\begin{tabular}{lrrllll}
\hline
\mc{1}{c}{$\beta$} & \mc{1}{c}{$L$} & \mc{1}{c}{stat$/10^5$}&\mc{1}{c}{$\chi$} 
 &\mc{1}{c}{$\xi_{exp}$} & \mc{1}{c}{$\xi_{2nd}$} & 
 \mc{1}{c}{$c_a \xi_{exp}/\xi_{2nd}$} \\
\hline
0.35544347& 48 & 150\phantom{00} & \phantom{00}10.15694(17) & \phantom{0}1.923431(29) &\phantom{0}1.944827(36)& 1.0000705(45) \\ 
0.36544347&64 &112\phantom{00} & \phantom{00}16.00107(29) & \phantom{0}2.456110(40) &\phantom{0}2.473081(49)& 1.0000823(43) \\
0.37544347&88 &107\phantom{00} & \phantom{00}33.28360(75) & \phantom{0}3.611554(68) &\phantom{0}3.623445(79)& 1.0001157(40) \\
0.37944347&112 &105\phantom{00} & \phantom{00}54.0942(13)  &\phantom{0}4.647652(90) &\phantom{0}4.65721(10) & 1.0001348(40) \\
0.38144347&132 & 111\phantom{00} & \phantom{00}76.0765(19)  &\phantom{0}5.54245(11)  &\phantom{0}5.55076(13) & 1.0001459(41) \\
0.38244347&148 & 101\phantom{00} & \phantom{00}94.2427(25)  &\phantom{0}6.18851(13)  &\phantom{0}6.19619(15) & 1.0001546(40) \\
0.38344347&168 & 101\phantom{00} & \phantom{0}122.1430(34) &\phantom{0}7.07068(15)  &\phantom{0}7.07771(18) & 1.0001625(42) \\
0.38444347&196 & 101\phantom{00}&  \phantom{0}169.6870(59) &\phantom{0}8.36931(19)  &\phantom{0}8.37571(22) & 1.0001700(45) \\
0.38544347&248 & 80\phantom{00}& \phantom{0}266.0115(91) &10.53510(27) &10.54096(30) & 1.0001809(47) \\
0.38644347&400 & 28\phantom{00} & \phantom{0}543.398(29)  &15.17511(58) &15.18080(65) & 1.0001944(66) \\ 
0.38714347&580 &12\phantom{00}&1449.01(15)   &25.0266(18)  &25.0332(20)  & 1.000199(12)  \\
\hline
\end{tabular}
\end{center}
\end{table}

We analysed the data for the second moment correlation length given here
along with those of ref. \cite{MHcorrections}. We used Ans\"atze of the type
\begin{equation}
 \xi_{2nd} = f_{2nd,+} t^{-\nu} 
\;\;\;\left( 1 + \sum_i^n a_i t^{\epsilon_i} \right)
\end{equation}
with $n=2$ and $3$. The reduced temperature is $t=\beta_c - \beta$.
Free parameters are $f_{2nd,+}$ and $a_i$. We fixed
$\nu = 0.629977$ and the correction exponents
$\epsilon_1=$ $0.832 \nu$, $\epsilon_2=1$ and $\epsilon_3 = \gamma \approx 
2 \nu$. Taking into account the results of various fits we arrive at
\begin{equation}
 f_{2nd,+} = 0.2284(1) - 2.1 \times (\nu - 0.629977) + 500 \times 
 (\beta_c - 0.387721735) \;\;.
\end{equation}
In a similar way we arrive at the estimate of the amplitude of the 
magnetic susceptibility
\begin{equation}
C_+ = 0.14300(5) - 1.2  \times (\gamma - 1.237084)
+ 300 \times (\beta_c - 0.387721735) \;\;.
\end{equation}

Next we studied amplitude ratios that combine the high with the low 
temperature phase.
Following \cite{CaHa97,MHamplitude}
we computed the ratios $R_{\chi}(\beta_l-0.387721735) = 
\chi(2 \times 0.387721735  - \beta_l)/\chi(\beta_l)$  and
$R_{\xi_{2nd}}(\beta_l-0.387721735)=
\xi_{2nd}(2 \times 0.387721735-\beta_l)/\xi_{2nd}(\beta_l)$. This way, 
the divergence is cancelled and the value of the critical exponent is 
not needed. We fitted these two quantities with the Ans\"atze 
\begin{equation}
\label{ansatzR2}
R(t) = R^* + a_1 t^{\epsilon_1} + a_2 t 
\end{equation}
and
\begin{equation}
\label{ansatzR3}
R(t) = R^* + a_1 t^{\epsilon_1} + a_2 t + c t^{\epsilon_3} \;\;,
\end{equation}
where we take $\epsilon_1=\nu \omega$ and $\epsilon_3=\gamma \approx 2 \nu$. 
In order to obtain the dependence of our result  on the value of $\beta_c$, 
we repeated the analysis, assuming $\beta_c=0.3877276$, which is our 
central estimate of $\beta_c$ plus the error bar. 
Our final estimates are
\begin{equation}
 \frac{C_{+}}{C_{-}} = 4.714(4) + 36000 \times (\beta_c - 0.387721735)
\end{equation}
and
\begin{equation}
 \frac{\xi_{2nd,+}}{\xi_{2nd,-}} = 1.940(2) + 11000 \times (\beta_c -0.387721735) \;.
\end{equation}
These results are consistent with $\frac{C_{+}}{C_{-}}=4.713(7)$ and
$\frac{\xi_{2nd,+}}{\xi_{2nd,-}}=1.939(5)$ given in \cite{MHamplitude}.  
For a detailed comparison with estimates obtained in the literature see 
\cite{MHamplitude}.

To get the universal amplitude ratio $Q_2$ we first analysed
\begin{equation}
r = \chi/\xi_{2nd}^{2-\eta}
\end{equation}
both for the high temperature phase as well as the critical isotherm.
We fitted our data with the Ansatz
\begin{equation}
r = r_{\infty} + a_1 \xi_{2nd}^{-\epsilon_1} + a_2 \xi_{2nd}^{-\epsilon_2}
\;\;,
\end{equation}
where $r_{\infty}$, $a_1$ and $a_2$ are the free parameters of the fit.
We fixed $\epsilon_1=0.832$ and $\epsilon_2=1.67$ or $2$. 
In the case of the high temperature phase the fits with $\epsilon_2=1.67$
are clearly better than those with $\epsilon_2=2$. Comparing the results
of different fits we arrive at $r_{\infty,high}=2.5960(15)$ for the 
high temperature phase. Here we have also taken into account the uncertainty
of $\eta$. In the case of the critical isotherm we arrive at 
$r_{\infty,c}=2.2020(20)$. As our result for the universal amplitude ratio 
we quote
\begin{equation}
Q_2= r_{\infty,high}/r_{\infty,c}=1.179(2) \;\;.
\end{equation}
This can be compared with $Q_2=  $ and $1.195(10)$ obtained in 
refs. \cite{EnFrSe03} and \cite{pisaseries}, respectively. For a 
comprehensive collection of results obtained by various methods, see
the tables 11 and 12 of the review \cite{PeVi}.

Finally, using the amplitudes computed above 
\begin{equation}
R_{\chi} = C_+ D_c B^{\delta-1} = 1.650(3)  \;\;.
\end{equation}
This result can be compared with $R_{\chi} = 1.723(13)$ obtained 
from Monte Carlo simulations of the improved $\phi^4$ model on 
the simple cubic lattice \cite{EnFrSe03}, and $R_{\chi} = 1.660(4)$ using 
high temperature series expansions of improved lattice models in 
combination with a parametric representation of the equation of state 
\cite{pisaseries}.  For results obtained by other methods see table 12
of the review \cite{PeVi}.

\section{Conclusions and Outlook}
We discuss a variance reduced estimator of the connected two-point function
that is based on the exchange cluster algorithm 
\cite{ReMaCh98,ChMaRe98,HeBl98}.
We studied the properties of this estimator at the example of the improved 
Blume-Capel model on the simple cubic lattice. We performed simulations
for the high and the low temperature phase at a vanishing external field 
and for the critical isotherm.
In the high temperature phase, we find that the variance reduced estimator
of the slice-slice correlation function $\bar{G}(r)$ based on the standard
Swendsen-Wang algorithm \cite{SwWa87} and on the Swendsen-Wang version of the 
exchange cluster algorithm perform similarly. In both cases, the relative
statistical error increases as $\exp(r/[2 \xi_{exp}])$.  This is a clear
improvement compared with $\exp(r/\xi_{exp})$ for the standard estimator.
The exchange cluster improved estimator still works in the presence of
a broken  $\mathbb{Z}_2$ symmetry.  For the critical isotherm as well 
as the low temperature phase we find that the relative statistical error
increases as $\exp(r/[2 \xi_{exp}])$ as it is the case in the high 
temperature phase. 
Analysing the slice-slice correlation function we confirm that for the 
low temperature phase, there is a second isolated exponentially decaying
term with $\xi_2 \approx \xi_{exp}/1.83$ \cite{CaHaPr99,CaHaPrZa02}. 
In contrast, for the critical isotherm, we do not find such a contribution.
The reduced statistical error allowed us to take the effective correlation 
length at a large separation of the slices as estimate of the exponential 
correlation length $\xi_{exp}$, reducing systematical errors to
one eighth of a per mille or less. This allows us compute
the ratio $f_{\exp}/f_{2nd}$ of the amplitudes of the exponential and the 
second moment correlation length with high precision. Using our data for
the magnetisation, the magnetic susceptibility and the correlation 
length, we computed various universal amplitude ratios. We compared our
estimates with those of refs. \cite{pisaseries,EnFrSe03,MHamplitude}. 
For a comprehensive review  of results obtained by various
methods see ref. \cite{PeVi}. 

It seems plausible that the variance reduced estimator discussed here is 
also effective for other models with $\mathbb{Z}_2$ symmetry.
However it is quite unclear how the idea can be generalized to problems with
an other symmetry. 
In our assessement, the main virtue of the exchange cluster
algorithm is the construction of variance reduced estimators of excess
quantities related to defects of various kinds in Ising-like systems. 
In refs. \cite{mysphere,myfilm} we computed the thermodynamic Casimir force  
using such an estimator.

\section{Acknowledgement}
This work was supported by the DFG under the grant No HA 3150/3-1.

\appendix

\section{The Ising model on the critical isotherm}
We simulated the Ising model at $\beta=0.22165462$, which is the estimate of 
the inverse critical temperature given in eq.~(A2) of \cite{MHcorr}.
We performed these simulations before we became aware
of the variance reduced estimators discussed in the main body
of the text. 
We simulated lattices with $L_0 > L=L_1=L_2$. Therefore we computed
the slice-slice correlation function in $0$-direction only. Also the ratio
$L/\xi$ is smaller than in our study of the improved Blume-Capel model.
However $L$ is large enough to ignore deviations from the thermodynamic limit.
Our results for the energy density $E=\frac{1}{L_0 L^2} 
\langle \sum_{<xy>} s_x s_y \rangle$, the magnetisation,
the magnetic susceptibility, the second moment correlation length and 
the dimensionless quantity $u$ are summarized in table \ref{finitehI}.
All estimates given here are computed for $R \approx 4 \xi_{eff}$, 
eq.~(\ref{extrapol}). Therefore we do not quote an estimate of $\xi_{exp}$.

\begin{table}
\caption{\sl \label{finitehI}
Results for the critical isotherm $\beta=0.22165462$ of the standard Ising model
on the simple cubic lattice. For a discussion see the text.
}
\begin{center}
\begin{tabular}{lcrlllll}
\hline
\mc{1}{c}{$h$} & \mc{1}{c}{$L_0 \times L^2$} & \mc{1}{c}{stat$/10^6$} & 
\mc{1}{c}{$E$} & \mc{1}{c}{$m$} & \mc{1}{c}{$\chi$}  & \mc{1}{c}{$\xi_{2nd}$} &
  \mc{1}{c}{$u$} \\
\hline
0.05   &  $32 \times 12^2$ & 200\phantom{00}  & 1.6576621(58) &  0.6819794(16) & \phantom{00}2.32857(15)  & 0.83556(24) & 25.748(21) \\
0.02   &  $48 \times 20^2$ & 200\phantom{00}  & 1.4119028(37) &  0.5794070(14) & \phantom{00}5.33012(35)  & 1.25521(37) & 24.085(20) \\
0.01   &  $64 \times 24^2$ & 200\phantom{00}  & 1.2833057(31) &  0.5087991(15) & \phantom{00}9.69650(64)  & 1.68910(41) & 23.318(16) \\
0.005  & $100 \times 36^2$ & 113\phantom{00}  & 1.1921108(26) &  0.4450808(16) & \phantom{0}17.3802(15) &  2.26080(81) & 22.778(23) \\
0.002  & $160 \times 50^2$ &  59\phantom{00}  & 1.1124843(25) &  0.3714069(21) & \phantom{0}37.0357(45) &  3.30542(16) & 22.303(30) \\
0.001  & $160 \times 68^2$ &  45\phantom{00}  & 1.0735366(24) &  0.3231855(26) & \phantom{0}65.174(11) &  4.4007(28) & 21.965(39) \\
0.0006 & $200 \times 82^2$  & 31\phantom{00}  & 1.0529398(24) &  0.2914486(31) & \phantom{0}98.538(21) &  5.4172(42) & 21.892(48) \\
0.00033& $300 \times 100^2$  & 27\phantom{00}  & 1.0351784(19)&  0.2580626(30) &159.599(34) &  6.9184(55) & 21.712(48) \\
\hline
\end{tabular}
\end{center}
\end{table}

We fitted the data with similar Ans\"atze as those for the improved 
Blume-Capel model in the main body of the text.
We fitted the magnetisation with Ans\"atze of the form
\begin{equation}
\label{fitm0}
 m = B_c \; h^{1/\delta} \;\left(1 + \sum_{i=1}^n a_i h^{\epsilon_i} \right)
\end{equation}
where we fixed $1/\delta$ and the correction exponents $\epsilon_i$.
The free parameters of the fit are $B_c$ and $a_i$. We used 
$\epsilon_1=0.832 \nu_c$,  $\epsilon_2=1.664  \nu_c$  or $2 \nu_c$ and 
$\epsilon_3=1$. 
For the Ansatz with $n=1$ we get an acceptable $\chi^2/$d.o.f. only when 
discarding most of the data points.
Taking into account $h=0.001$, $0.0006$ and $0.00033$
we get $\chi^2/$d.o.f. $=1.56$, $B_c=1.395500(55)$ and  $a_1=-0.2066(4)$. 
Next we performed fits with $n=2$ correction terms.
Among our different choices the smallest $\chi^2/$d.o.f. are found 
for $\epsilon_2=1.664 \nu_c$. 
Here we get, taking $h=0.01$ down to $0.00033$ into account,
$\chi^2/$d.o.f. $=1.40$, 
$B_c=1.394070(34)$, $a_1=-0.1825(3)$ and $a_2=-0.1413(11)$.
Finally for $n=3$, with $\epsilon_2=1.664 \nu_c$ and $\epsilon_3=1$,
we get $\chi^2/$d.o.f. $=1.40$ 
taking all values of $h$. The results for the free parameters of the fit are
$B_c=1.393971(39)$, $a_1=-0.1806(5)$, $a_2=-0.166(4)$  and $a_3=0.043(6)$.

Based on these fits we arrive at our final estimate
\begin{equation}
 B_c = 1.3941(6) \;\;,\;\;\;  a_1 = -0.19(2) \; .
\end{equation}

Performing similar fits, we arrive at 
\begin{equation}
f_{2nd,c} = 0.2771(12)
\end{equation}
for the amplitude of the second moment correlation length.
We analysed our data for the coupling $u$ by fitting with the
Ansatz~(\ref{ucansatz}). We arrive at
\begin{equation}
 u_c = 21.05(15)  \;\;,
\end{equation}
which is consistent with our result~(\ref{finaluc}) obtained from the data 
for the improved Blume-Capel model.

\end{document}